\documentclass[reprint,amsmath,amssymb,aps,pra,]{revtex4-1}
\usepackage{graphicx}
\usepackage{subfig}
\usepackage{dcolumn}
\usepackage{bm}
\usepackage{soul}
\usepackage{color}

\begin{document}

\preprint{APS/123-QED}

\title{Binary Discrimination in Quantum Systems via Hypothesis Testing}

\author{Beili Gong}
\author{Wei Cui}
\email{aucuiwei@scut.edu.cn}
\affiliation{School of Automation Science and Engineering, South China University of Technology, Guangzhou 510641, China
}


\date{\today}

\begin{abstract}
We investigate the discrimination of two candidates of an unknown parameter in quantum systems with continuous weak measurement, inspired by the application of hypothesis testing in distinguishing two Hamiltonians [Kiilerich and M{\o}lmer, Phys. Rev. A, 98, 022103 (2018)].
Based on the measurement output and stochastic master equation, temporal evolutions of posterior probabilities of two hypotheses are given by Bayes' formula.
The Bayes criterion is presented by the likelihood ratio conditioned on the outcome of measurements.
Different from the calculation method based on maximum a posteriori criterion, the Bayes criterion based method for calculating the average probability of making errors is more suitable and efficient in general situation of binary discrimination.
Finally, an example of distinguishing two candidate Hamiltonians is given and the running times of calculating the average probability of error under the Bayes criterion and the maximum a posteriori criterion are compared to illustrate the feasibility of the hypothesis testing in quickly distinguishing two candidates of the parameter to be estimated.
\end{abstract}

\maketitle

\section{Introduction}
Parameter estimation refers to the procedure of using measurement output to estimate the unknown parameters, which has caused a wide range of interests in quantum systems \cite{Iwasawa:2013,Cortez:2017,Laverick:2018,ang2013optomechanical}.
Since, in most quantum systems, the system state can be indirectly measured in general and classical estimation theories are hard to apply to the measurement output directly, it technically difficult to construct an estimator for some unknown parameters.
However, selecting a value, which can best characterize the system dynamics, from some potential values as an estimator of an unknown parameter is an effective method in Hamiltonian estimation \cite{Ralph2000,Aharonov2002,Chase2009}.
From this method, how to distinguish the potential values in quantum systems based on measurement output becomes a key task.

There have already been many methods applied to discriminate possible quantities in physical systems, such as the
method based on the uncertainty relations \cite{Aharonov2002}, the Bayesian method \cite{lavine1992bayesian,Lin2015} and the kernel method \cite{AITCHISON1976}.
In recent years, a method of statistical inference named statistical hypothesis testing has been widely used to discriminate quantum states \cite{Yuen1975,Hiai2009,Kumagai:2013,Haikka:2017}, quantum channels \cite{Acin:2001,Chiribella:2008,Dumitrescu2016}, system models \cite{Tsang2012,turinici2003optimal}, quantum processes \cite{Wang2016,Klaus2015} and quantum Hamiltonians \cite{Klaus2015,Molmer2018,Kiilerich20182}.
According to the unnormalized stochastic master equations with different measurement schemes, such as photon counting, homodyne detection and an optimal projective measurement on its conditioned final state, and the corresponding measured datum, the distinction of two Hamiltonians has been achieved via hypothesis testing \cite{Molmer2018}.
Moreover, an upper bound, namely the average probability of error, for the ability to distinguish two potential Hamiltonians for a quantum circuit model has been derived \cite{Klaus2015} and to discriminate $N$ arbitrary quantum states for an open quantum system has been calculated by numerically efficient means \cite{Kiilerich20182}.
A classical framework of applying hypothesis testing to parameter estimation is first to specify a set of hypotheses of candidates of the unknown parameter, and then use a criterion to test the hypotheses according to the  observed data and finally accept one of the hypotheses.
In doing the test, it is of importance to study the average probability of error, i.e., the average probability of being assigned to erroneous hypothesis under a criterion.
Considering the maximum a posteriori criterion, the average probabilities of error for different measurement schemes have been calculated \cite{Molmer2018}.
However, the calculation methods depend on counting each judgment result and require a large number of trials, which results in increasing the computational complexity of calculating the average probability of error so that a significant amount of time  would be spent in parameters discrimination.
Furthermore, the aforementioned papers have focused on the parameters in Hamiltonian rather than other parameters in quantum systems.
Therefore, how to quickly distinguish potential values of an unknown parameter in quantum systems based on the measured data via hypothesis testing is still an open problem.

Note that the dynamic of a quantum system under continuous measurement can be described by a stochastic master equation \cite{Gammelmark2013,Molmer2018}.
Together with some prior knowledge of the parameters, we can construct some candidate systems with the same structure such that the hypothesis testing can be used.
This paper aims at distinguishing two candidates of an unknown parameter for a quantum system with continuous weak measurement by using hypothesis testing.
According to the measured data  and the Bayes' formula \cite{Kiilerich2016}, we obtain the posterior probabilities of two hypotheses, respectively.
Furthermore, the Bayes criterion is presented based on a likelihood ratio \cite{Burnham2002Model}, and an algorithm to calculate the average probability of error is proposed.
A numerical example of distinguishing two candidates of an unknown parameter in Hamiltonian is illustrated to show the feasibility of the proposed hypothesis testing method. The running times of calculating the average probability of error under the maximum a posteriori criterion and the Bayes criterion are also compared to illustrate the efficiency of proposed method.

This paper is organized as follows.
In Sec.~\ref{Sec:model}, a brief introduction of an quantum system and a reduced stochastic master equation are presented.
The measurement record via continuous weak measurement is exhibited as well.
In Sec.~\ref{Sec:estimation}, based on the Bayes' formula and Bayes criterion, the procedure of distinguishing two candidates using hypothesis testing is described in detail, and the average probability of error is given.
Some numerical simulations illustrate the feasibility and effectiveness of the proposed method in binary discrimination.
We summarize our conclusion in Sec.~\ref{Conclusion}.

\section{Model}\label{Sec:model}
Consider a two-level quantum system with the following Hamiltonian
\begin{eqnarray}\label{Hamiltonian}
 H=\frac{\Omega}{2} \sigma_x + \frac{\Delta}{2}\sigma_z,
\end{eqnarray}
where $\Omega$ and $\Delta$ are the detuning and the Rabi frequency \cite{You:2005,Devoret:2013,Yang:2018}, and $\sigma_x, \sigma_z$ are Pauli matrices.
Under continuous weak measurement, the evolution of the system state can be described by a reduced stochastic master equation \cite{Qi:2010,Cui:2013,Jacobs:2014}, i.e.,
\begin{eqnarray}\label{stochastic_master_equation}
d{\tilde \rho _t} =  - i[H,{\tilde \rho_t }]dt + \eta D[F]{\tilde \rho_t}dt+ \sqrt{\eta}\mathcal{M}[F]{\tilde \rho_t }d{Y_t},
\end{eqnarray}
where $\mathcal{D}[F]\rho = F\rho F^{\dag}-\frac{1}{2}(FF^{\dag} \rho+\rho FF^{\dag})$ and $\mathcal{M}[F]\rho = (F{\rho } + \rho F^\dag)$.
Here, $\tilde \rho_t$ is the unnormalized state, $\eta$ and $F$ are the measurement efficiencies and measurement operator, respectively, and $d{Y_t}$ describes the quantum fluctuations of the continuous output signal, which is given by
\begin{eqnarray}\label{Eq:measurement_output}
 dY_t = \sqrt{\eta}{\rm{Tr}}(\mathcal{M}[F]\rho_t)dt + dW_t.
\end{eqnarray}
Here $dW_t$ is the Wiener increment with zero mean and variance $dt$, and $\rho_t$ is the normalized quantum state \cite{Gammelmark2013,Wiseman2009} satisfying
\begin{eqnarray}\label{Eq:relationship}
{\rho _t} = \frac{{{{\tilde \rho }_t}}}{{{\rm{Tr}}({{\tilde \rho }_t}})}.
\end{eqnarray}

By substituting Eqs.~\eqref{Eq:measurement_output} and \eqref{Eq:relationship} into Eq.~\eqref{stochastic_master_equation}, the normalized quantum stochastic master equation can be written as
\begin{eqnarray}\label{Eq:stochastic_master_equation1}
d{\rho _t} =  - i[H,{\rho _t}]dt + \eta \mathcal{D}[F]{\rho _t}dt + \sqrt{\eta}\mathcal{H}[F]\rho_td{W_t},
\end{eqnarray}
with
$\mathcal{H}[F]\rho = \mathcal{M}[F]\rho-{\rm{Tr}}(\mathcal{M}[F]\rho)\rho$.
A detailed derivation of the above stochastic master equation was presented in our previous work \cite{Gong:2018}.
From Eq.~\eqref{Eq:stochastic_master_equation1}, one can intuitively see that the evolutions of unnormalized quantum states can be characterized by the evolutions corresponding to normalized states.

\section{Binary Discrimination and Computational Complexity}\label{Sec:estimation}
Parameters describe an underlying physical setting in such a way that their value affects the distribution of the measurement output.
Generally, it is much easier to derive an estimator of the unknown parameter in a classical system from limited information than that in a quantum system.
In other words, obtaining an explicit estimator of a unknown parameter in a quantum system is nearly impossible.
On the other hand, if some potential values of the unknown parameter are known \emph{a prior}, then it is natural to select the most proper value as the estimate of the parameter.
In this work we will consider distinguishing the potential values of an known parameter and select an appropriate value via hypothesis testing.

Without loss of generality we assume there are two candidates of a parameter to be estimated in a quantum system.
Correspondingly, two candidate quantum systems are constructed based on the stochastic master equation \eqref{Eq:stochastic_master_equation1} and the measurement output.
For determining which hypothetical system is more suitable to describe the observed system, a Bayes criterion is proposed and the average probabilities of making error decision is calculated.
This section consists of three parts: introducing the hypothesis testing, illustrating the process of distinction by an example, and comparing the computational complexities of calculating the average probabilities of error between using maximum a posteriori criterion and using Bayes criterion.

\subsection{Hypothesis testing}
\subsubsection{Two hypotheses and posterior probability}
Let $\theta$ be any unknown parameter to be estimated in a quantum system, which might be Rabi frequency, detuning, dissipation rates, measurement strength, and so on.
Suppose $\theta_0$ and $\theta_1$ are two candidate values closest to the unknown parameter and the corresponding hypotheses are $H_0:\theta = \theta_0$ and $H_1:\theta = \theta_1$, respectively.
We denote the prior probabilities of the two hypotheses by $P({H_0})$ and $P({H_1})$, and the measurement output by $D_t$.
Note that Bayes' rule is a celebrated formula tells that how to update the probabilities of hypotheses as more information becomes available.
Accordingly, the posterior probability $P({H_i|{D_t}})~(i=0,1)$, which is referred to as the conditional probability of accepting $H_i$ when observing $D_t$, can be expressed as
\begin{eqnarray}\label{Eq:Bayes rule}
P({H_i}|{D_t}) = \frac{{P({D_t}| {{H_i}})P({H_i})}}{P({D_t})},
\end{eqnarray}
where $P( {D_t|{H_i}})$ is the probability of obtaining $D_t$ given the hypothesis $H_i$, and $P({D_t})= {{\sum_{j = 0}^1{P({D_t}|{{H_j}})P({H_j})}}}$ is the marginal probability for all possible hypotheses being considered .

Define $L\left( D_t| H_i \right)=P\left( D_t| H_i \right)/P_0(D_t)$ as a likelihood function when assuming $H_i$ is true, where $P_0(D_t)$ is the probability density for a Poisson or Wiener process \cite{Gammelmark2013}.
According to the trajectory of unnormalized state, the likelihood function of a specific sequence of detection events is simply $L(D_t|H_i) = {\rm{Tr}}(\tilde{\rho}_t|{H_i},{D_t})$ given hypothesis $H_i$, and then the posterior probability $P\left( {H_i\left| {D_t}\right.} \right)$ can be further written as
\begin{eqnarray}\nonumber\label{Eq:Bayes rule1}
P({H_i}\left| {{D_t}} \right.) = \frac{{{\rm{Tr}}({{\tilde \rho_t }}\left| {{H_i},{D_t}} \right.)P({H_i})}}{{\sum_{j = 0,1} {\rm{Tr}}({{\tilde \rho_t }}\left| {{H_j},{D_t}} \right.)P({H_j})} }.
\end{eqnarray}
Together with Eq.~\eqref{Eq:stochastic_master_equation1}, the derivative of the likelihood function $L\left( D_t| H_i \right)$ with respect to $t$ is given by
\begin{eqnarray}
d{L\left( D_t| H_i \right)} &= {\rm Tr}\left( {d{{\tilde \rho }_t}} \right) = \sqrt {\eta} {\rm Tr}\left( {\mathcal{M}\left( {{{\tilde \rho}_t}} \right)} \right)d{Y_t} \nonumber\\
&= \sqrt {\eta} {\rm Tr}\left( {\mathcal{M}\left( {{\rho _t}} \right)} \right){L\left( D_t| H_i \right)}dY_t. \nonumber
\end{eqnarray}
Define $l_t\left( D_t| H_i \right)=  {\ln} L\left( D_t| H_i \right)$ as the logarithmic likelihood.
Then the derivative of the logarithmic likelihood is given by
\begin{eqnarray}\label{Eq:l_t}
dl\left( D_t| H_i \right)=  \sqrt{\eta}{\rm{Tr}}(\mathcal{M}[F]\rho_t)dY_t.
\end{eqnarray}
It is clear that the evolution of the logarithmic likelihood function only depends on the measurement output and the normalized stochastic master equation so that one can obtain it from combining Eqs.~\eqref{Eq:stochastic_master_equation1} and \eqref{Eq:l_t}.

\subsubsection{Bayes criterion}
To determine the most probable hypothesis, it is natural to compare the observation probabilities conditioned on assumed hypotheses.
A simplest decision rule of whether accept the hypothesis or not is that accept $H_i$ if $P(H_i|D_t)> P(H_{\{0,1\}\backslash i}|D_t),~i=0,1$, which can be standardly expressed as
\begin{eqnarray}\label{Eq_decision1}
{P({H_0}|{D_t})}\mathop{\gtrless}\limits_{{H_1}}^{{H_0}} {P({H_1}|{D_t})}.
\end{eqnarray}
Substituting Eq.~\eqref{Eq:Bayes rule} into the decision rule \eqref{Eq_decision1}, we have
\begin{eqnarray}\label{Eq_decision2}
\frac{{P({D_t}\left| {{H_0}} \right.)P({H_0})}}{P\left( {D_t} \right)} \mathop{\gtrless}\limits_{{H_1}}^{{H_0}} \frac{{P({D_t}\left| {{H_0}} \right.)P({H_1})}}{P\left( {D_t} \right)}.
\end{eqnarray}
In order to quantify the performance of the decision rule, we define the Bayes risk as the expected value of the cost function
\begin{eqnarray*}
E[{C}] = \sum_{i = 0}^1 {\sum_{j = 0}^1 {{C_{ij}}} } P({H_i}\left| {{H_j}} \right.)P({H_j}),
\end{eqnarray*}
where $C_{ij}$ and $P({H_i}\left| {{H_j}} \right.)$ respectively represent the cost function of making wrong decision that accept hypothesis $H_i$ while $H_j$ is true and the corresponding probability.
Through minimizing the Bayes risk \cite{Hoballah:1989,Brandt:2014}, we can easily obtain the Bayes criterion
\begin{eqnarray}\label{eq:final_rule}
{\Lambda } = \frac{P(D_t|H_0)}{P(D_t|H_1)} \mathop{\gtrless}\limits_{{H_1}}^{{H_0}} \frac{({C_{01}-C_{11}})}{({C_{10}-C_{00}})}\alpha ,
\end{eqnarray}
where $\alpha = P({H_1})/P({H_0})$.
Here, $\Lambda$ is referred to as the likelihood ratio.
Note that the cost functions should satisfy ${C_{01} > C_{11}}$ and ${C_{10} > C_{00}}$.
For simplicity, applying the zero-one cost function
\begin{equation*}
 {C_{ij}} = \left\{ \begin{array}{ll}
0,&i = j\\
1,&i \ne j
\end{array} \right.
\end{equation*}
and taking the logarithm of both sides of Eq.~\eqref{eq:final_rule}, we obtain
\begin{eqnarray}\label{Eq:Bayes_criterion}
\ln \Lambda \mathop{\gtrless}\limits_{{H_1}}^{{H_0}} \ln \alpha.
\end{eqnarray}
That is, if $\ln \Lambda > \ln \alpha$, the hypothesis $H_0$ would be accepted so that $\theta_0$ is selected as an estimate of the unknown parameter; otherwise accept $H_1$ and select $\theta_1$.
It is observed that the simplified Bayes criterion \eqref{Eq:Bayes_criterion} only depends on the normalized quantum stochastic master equation \eqref{Eq:stochastic_master_equation1} and Eq.~\eqref{Eq:l_t}.

\begin{figure}[hbt]
\centering
\includegraphics[width=3.4in]{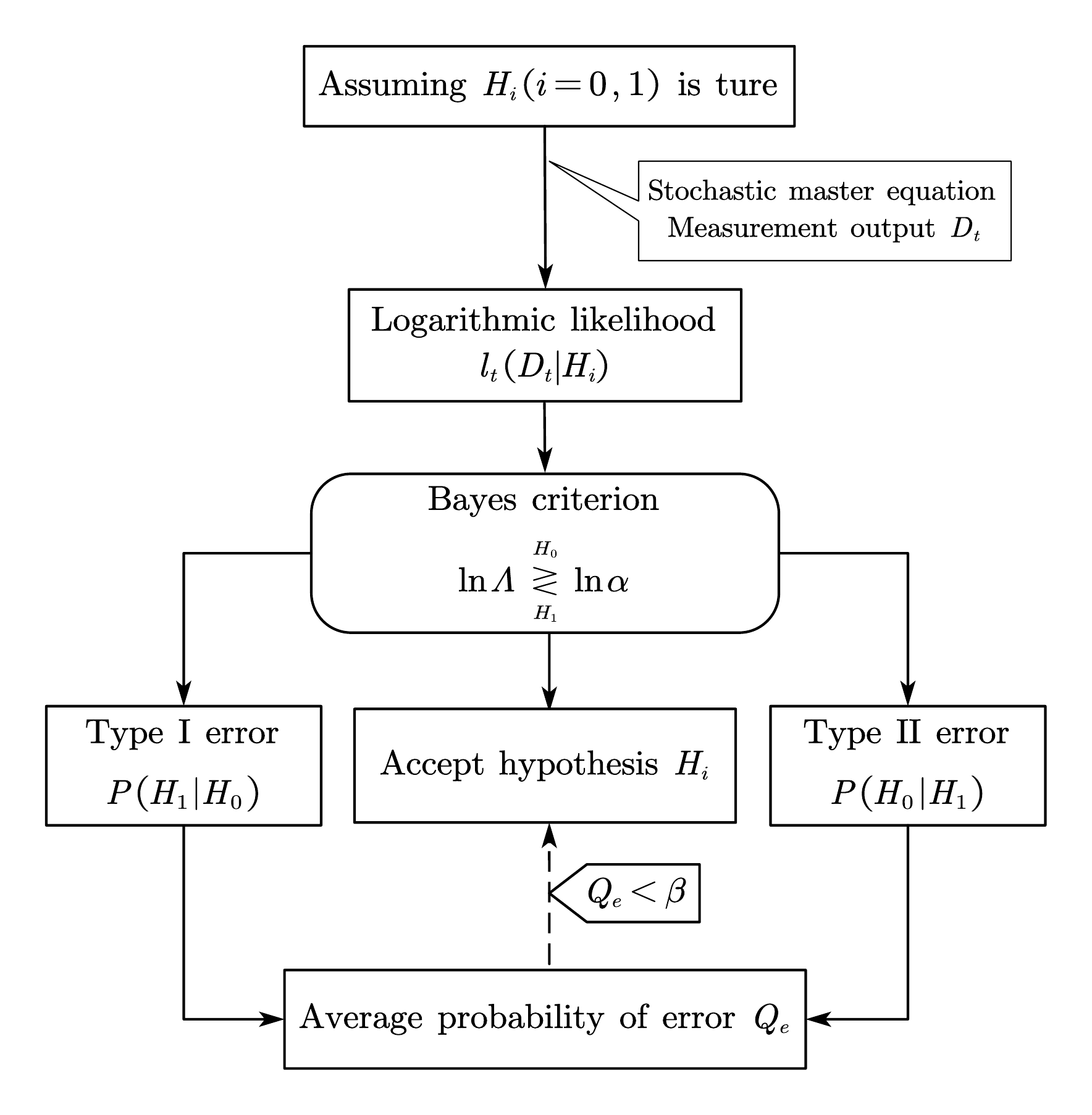}
\caption{Flowchart of binary discrimination via hypothesis testing.} \label{Fig:chart}
\end{figure}

\begin{figure}[hbt]
\begin{minipage}[h]{1\linewidth}
\centering
\includegraphics[width=3in]{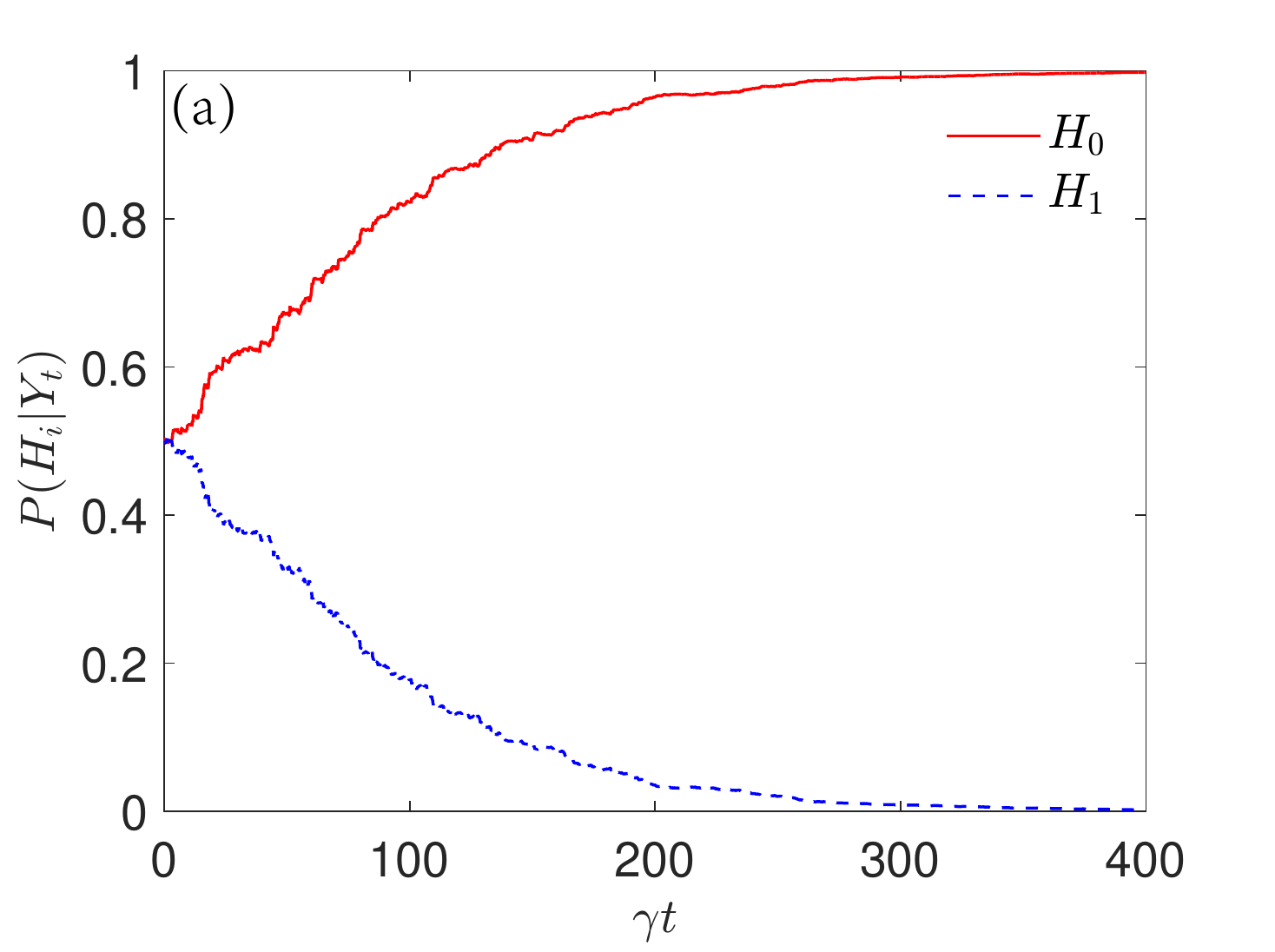}
\end{minipage}

\begin{minipage}[h]{1\linewidth}
\centering
\includegraphics[width=3in]{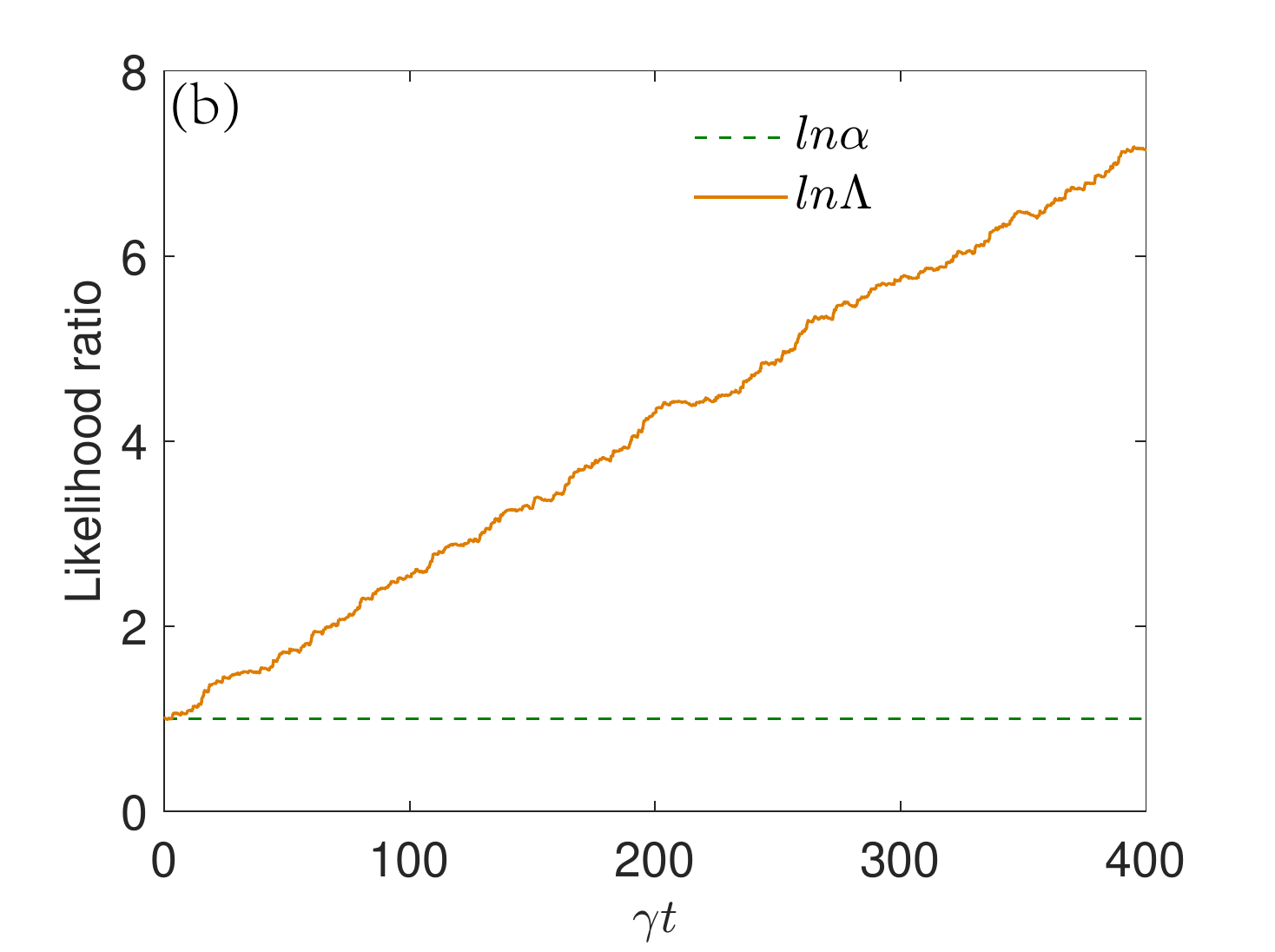}
\end{minipage}
\caption{Temporal evolutions of (a) the posterior probability $P({H_i}\left| {{Y_t}} \right.)$  and (b) the likelihood ratio  for a single measurement when $H_0$ is true.}
\label{Fig:figs1}
\end{figure}

\subsubsection{Average probability of error}
The Bayes criterion \eqref{Eq:Bayes_criterion} provides a simple way to accept one of the two hypotheses or, equivalently, select one of the two candidates of unknown parameters.
Even so, it is possible for the criterion to make wrong decisions or error.
Meanwhile, there are two types of error: type I error and type II error.
A type I error occurs when $H_0$ is true but is rejected (accept $H_1$), and a type II error occurs when $H_0$ is false but erroneously fails to be rejected.
As a result, we define the average probability of assigning an erroneous hypothesis 
\begin{eqnarray}\label{Eq:error_probability}
{Q_e} = P({H_1}\left| {{H_0}} \right.)P({H_0}) + P({H_0}\left| {{H_1}} \right.)P({H_1}),
\end{eqnarray}
where $P({H_1}\left| {{H_0}} \right.)$ and $P({H_0}\left| {{H_1}} \right.)$ respectively denote the probabilities of type I error and type II error, and $P({H_0})$ and $P({H_1})$ are respectively the prior probabilities of the corresponding hypotheses.
Taking Eq.~\eqref{Eq:Bayes_criterion} into consideration, it is clear that minimizing the Bayes risk actually corresponds to minimize the average probability of error.
Therefore, the effectiveness of the Bayes criterion can be illustrated by calculating the average probability of error.

For convenience, we define $\beta$ as an error threshold.
That is, when ${Q_e}<\beta$ we can stop observing the system and say that the estimation of the unknown parameter is done, i.e., the hypothesis is accepted or rejected.
Note that the smaller $\beta$ is, the more confident the decision-making is.
Based on the measured data and the proposed Bayes criterion, the procedure of distinguishing two candidates of an unknown parameter via hypothesis testing is shown in Fig.~\ref{Fig:chart}.

\subsection{Numerical example}\label{Sec:simulation}

To illustrate the feasibility and effectiveness of the proposed hypothesis testing in discrimination of candidate values of an unknown parameter, we consider the estimation of an unknown parameter in the system Hamiltonian.
Let $\Omega$ in Eq.~\eqref{Hamiltonian} be the unknown parameter that requires estimating, and the corresponding hypotheses are $H_0:\Omega = \gamma$ and $H_1:\Omega = 2\gamma$ with prior probabilities $P(H_0)=P(H_1)=0.5$.
The other parameters of the system are given by $\Delta = 1.43\gamma$, $\eta=0.5$, and the measurement operator is $F=\sqrt{\kappa}\sigma_z$ with the measurement strength $\kappa = 1$.
The initial state of the system is set to $x(0)=y(0)=0, z(0)=1$.

\begin{figure}[hbt]
\centering
\includegraphics[width=3in]{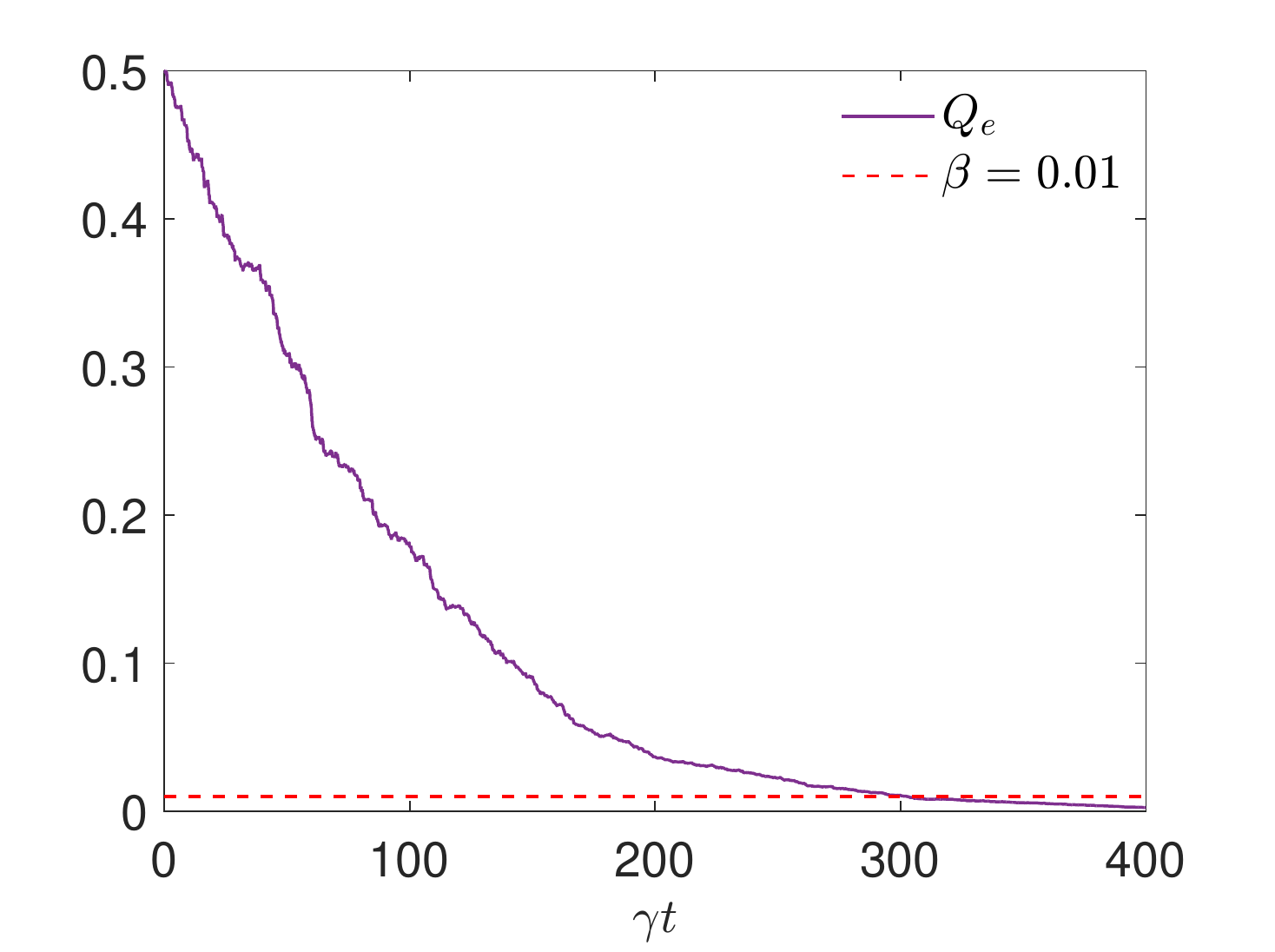}
\caption{\label{Fig:figs2} Temporal evolutions of the average error probability $Q_e$ for a single experiment.}
\end{figure}

Assuming $H_0$ is true, the evolution of posterior probability $P({H_0}\left| {{Y_t}} \right.)$ for a single measurement can be obtained based on the normalized quantum stochastic master equation, as shown in Fig.~\ref{Fig:figs1}(a).
The red solid curve is the trajectory of the conditional probability $P({H_0}\left| {{Y_t}} \right.)$ and the blue dashed one is the evolution of $P({H_1}\left| {{Y_t}} \right.) = 1-P({H_0}\left| {{Y_t}} \right.)$, which is referred to as the probability of making type I error.
Meanwhile, the orange solid curve in Fig.~\ref{Fig:figs1}(b) depicts the temporal evolution of logarithmic likelihood ratio under a single measurement.
This means that the hypothesis $H_0$ will be accepted, which confirms that the Bayes criterion is reliable.

According to Fig.~\ref{Fig:chart}, the trajectory of the average probability of error $Q_e$ is also plotted in Fig.~\ref{Fig:figs2} for a single experiment, where $\beta=0.01$.
It shows that the average probability of error would approach to $0$, which also reveals the feasibility of the proposed hypothesis testing in binary discrimination.

\subsection{Computational complexity}

In this part, the running times, which will be defined clearly, of calculating the average probability of error for the proposed algorithm and another algorithm based on the maximum a posteriori criterion, developed by Kiilerich and M{\o}lmer in \cite{Molmer2018}, are compared.
Note that the latter algorithm is a special case of Bayes criterion that the cost functions satisfy ${C_{01}- C_{11}}={C_{10} - C_{00}}$ and $\alpha = 1$.
Then the hypothesis $H_i$ is accepted when the posterior probability $P({H_i}\left| {{Y_t}} \right.)$ obtained by Eq.~\eqref{stochastic_master_equation} is larger than $1/2$.
Also, the probability of making error decision can be estimated by $P({H_i}|{H_j}) = {{n_i^{(j)}} \mathord{/
{\vphantom {{n_i^{(j)}} N}}\kern-\nulldelimiterspace} N}(i\ne j)$, where $n_i^{(j)}$ is the number of making errors in $N$ experiments.

\begin{table}[hbt]
\caption{Comparison of the running times of two algorithms when $\beta = 0.01$.}
\label{tab:example}
\vspace{-0.3cm}
\begin{ruledtabular}
\begin{tabular}{ccc}
$N$ & proposed algorithm &
\multicolumn{1}{c}
{ algorithm in \cite{Molmer2018}}\\
\colrule
1 & 2.4914s & ------ \\
10 & 24.1799s & 44.4146s \\
20 & 51.7167s & 88.3760s \\
50 & 131.8677s & 222.0794s \\
100 & 258.0833s & 445.1378s \\
\end{tabular}
\end{ruledtabular}
\end{table}

Based on the definition of  the error threshold, we consider the time taken for the average probability of error to reach $\beta$, namely the first-passage time, as the running time of an algorithm.
We implement the comparison work on PC ideacentre AI0 520-27ICB: a 2.4GHz Intel i7 CPU and 16GB memory.
Under the same setting of parameters in the above example, the first-passage times of the proposed algorithms and another one are depicted in Tab.~\ref{tab:example} when $\beta = 0.01$.
The smallness of the running time reveals the efficiency of the proposed algorithm, which means that the unknown parameter can be quickly determined from two candidates.

\section{Conclusion}\label{Conclusion}
In this paper, we have proposed a procedure of distinguishing two candidates of a unknown parameter in a quantum system by hypothesis testing.
Based on a normalized quantum stochastic master equation, we have updated the information about the parameter to be estimated by Bayes' formula together with the measured data, and obtained the posterior probability of two hypotheses.
Then we have built a Bayes criterion to accept or reject the hypotheses, and proposed an algorithm to calculate the average probability of making two types of erroneous decisions.
Finally, the feasibility and efficiency of hypothesis testing in binary discrimination has been illustrated by taking a Hamiltonian parameter as an example and comparing with the running times of calculating the average probability of error under the Bayes criterion and another criterion.

\begin{acknowledgements}
This work was supported by the National Natural Science Foundation of China under Grant 61873317 and in part by the Fundamental Research Funds for the Central Universities. 
\end{acknowledgements}

\end{document}